\documentclass[conference]{IEEEtran}

\usepackage{lmodern}  
\usepackage{mathptmx} 

\usepackage[T1]{fontenc}
\usepackage{mathtools}
\usepackage[inline]{enumitem}

\usepackage{mathtools}

\usepackage[nolist]{acronym}

\usepackage{graphicx}
\usepackage[tight,footnotesize]{subfigure}
\usepackage{epstopdf}

\usepackage{subcaption}  

\usepackage{xcolor}

\usepackage{makecell} 

\usepackage{url}
\usepackage{multirow}
\usepackage{balance}
\usepackage[numbers,sort&compress]{natbib}
\usepackage{amsmath}
\usepackage{amssymb}
\usepackage{bm}
\usepackage{lipsum}
\usepackage{color}
\usepackage{threeparttable}

\usepackage{physics} 

\usepackage{multirow}
\usepackage{authblk}

\usepackage[linesnumbered,ruled,vlined]{algorithm2e}

\SetCommentSty{mycommfont}

\SetKwInput{KwInput}{Input}                
\SetKwInput{KwOutput}{Output}              

\usepackage{lscape}
\usepackage{longtable}

\begin{document}

\title{Enhancing Open RAN Digital Twin Through Power Consumption Measurement}

\author{
Ahmed Al-Tahmeesschi\IEEEauthorrefmark{1}, 
Yi Chu\IEEEauthorrefmark{1}, 
Josh Shackleton, \\ 
Swarna Chetty, 
Mostafa Rahmani, 
David Grace 
and Hamed Ahmadi\
\\
School of Physics Engineering and Technology, University of York, United Kingdom\\
}

\maketitle
\IEEEpubidadjcol

\def\thefootnote{*}\footnotetext{These authors contributed equally to this work.}

\begin{abstract}
The increasing demand for high-speed, ultra-reliable and low-latency communications in 5G and beyond networks has led to a significant increase in power consumption, particularly within the Radio Access Network (RAN). This growing energy demand raises operational and sustainability challenges for mobile network operators, requiring novel solutions to enhance energy efficiency while maintaining Quality of Service (QoS). 5G networks are evolving towards disaggregated, programmable, and intelligent architectures, with Open Radio Access Network (O-RAN) spearheaded by the O-RAN Alliance, enabling greater flexibility, interoperability, and cost-effectiveness. However, this disaggregated approach introduces new complexities, especially in terms of power consumption across different network components, including Open \acp{RU}, Open \acp{DU} and Open \acp{CU}. Understanding the power efficiency of different O-RAN functional splits is crucial for optimising energy consumption and network sustainability. In this paper, we present a comprehensive measurement study of power consumption in RUs, DUs and CUs under varying network loads, specifically analysing the impact of \ac{PRB} utilisation in Split 8 and Split 7.2b. The measurements were conducted on both software-defined radio (SDR)-based RUs and commercial indoor and outdoor \ac{RU}, as well as their corresponding \ac{DU} and \ac{CU}. By evaluating real-world hardware deployments under different operational conditions, this study provides empirical insights into the power efficiency of various O-RAN configurations.
The results highlight that power consumption does not scale significantly with network load, suggesting that a large portion of energy consumption remains constant regardless of traffic demand. 

\end{abstract}

\begin{IEEEkeywords}
O-RAN, Open RAN, Energy Efficiency, Test
Methodology, Digital Twin, RU
\end{IEEEkeywords}
\IEEEpeerreviewmaketitle


\section{Introduction}
\label{sec:Introduction}
\begin{figure*}[!t]
    \centering
    \subfigure[Split 8]{%
        \includegraphics[clip, trim=0.0cm 7cm 0.0cm 3cm, width=1.75\columnwidth]{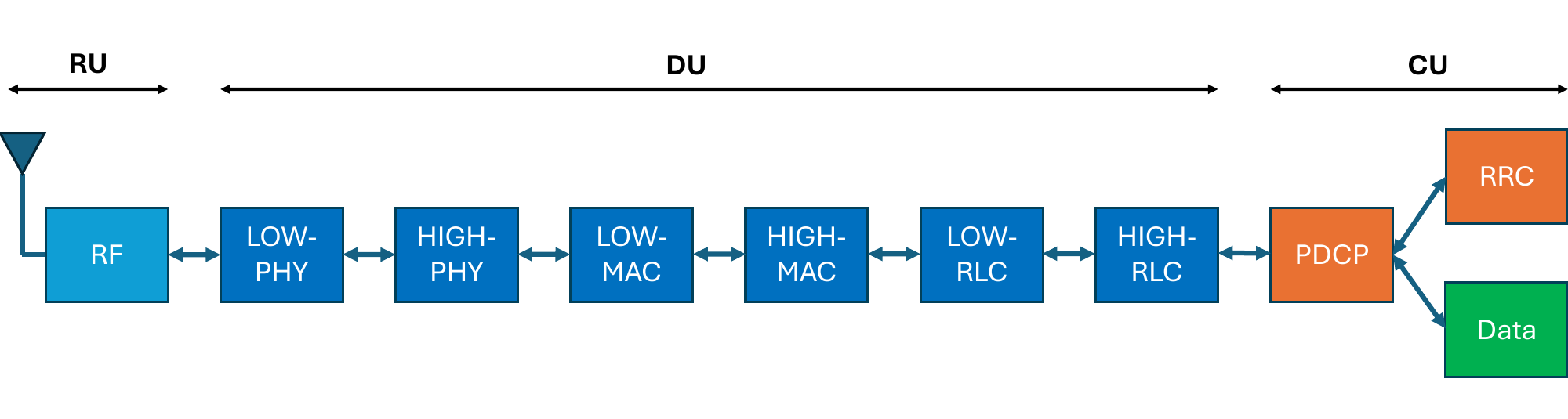}
        \label{fig:split_8}
    }
    \hfill
    \subfigure[Split 7.2b]{%
        \includegraphics[clip, trim=0.0cm 7cm 0.0cm 3cm, width=1.75\columnwidth]{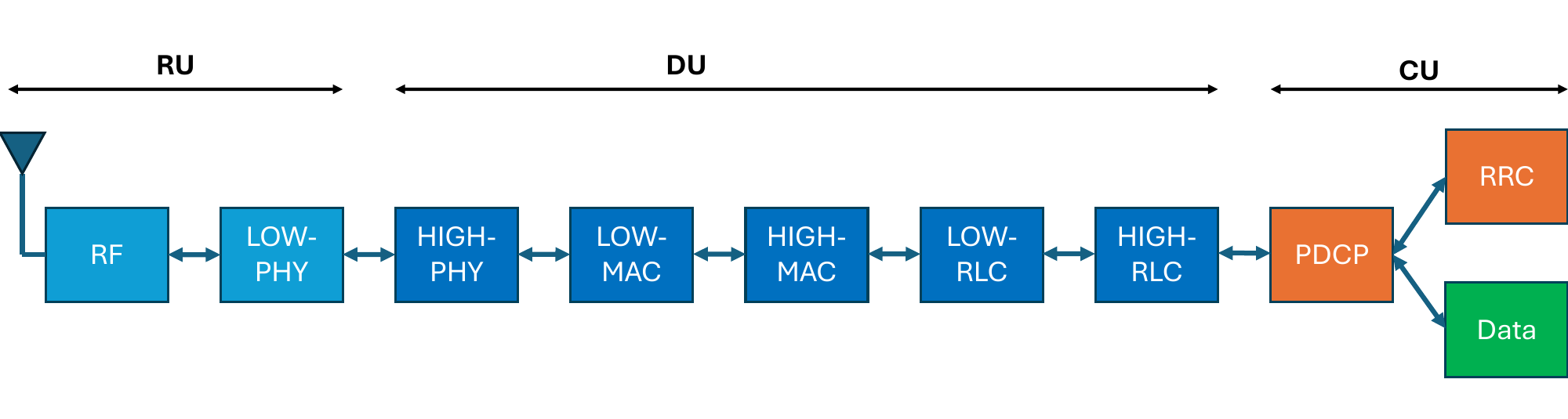}
        \label{fig:split_7}
    }
    \caption{Illustration of functional splits. \subref{fig:split_8} Split 8, \subref{fig:split_7} Split 7.2b.}
    \label{fig:locations_fig_1}	
    \vspace{-2mm}
\end{figure*}
The exponential growth in data demand, driven by emerging applications such as \acp{IC} \cite{IC_frontiers_survy_2023}, vehicle-to-everything (V2X) communication and other data-intensive use cases, has significantly escalated the power consumption of mobile networks \cite{Ericsson5G}. Next-generation mobile networks must support these computationally intensive applications while ensuring energy-efficient operation.

With \ac{5G} and beyond networks transitioning towards a more open, disaggregated, and virtualised approach to \ac{RAN} design due to its flexibility interoperability and potential cost efficiency \cite{ORAN_Architecture_2022} the Open \ac{RAN} paradigm, driven by the O-RAN Alliance, is poised to significantly impact RAN deployment and management. The disaggregation of the \ac{RAN} into distinct network functions, namely the \ac{CU}, \ac{DU} and \ac{RU}, introduces new opportunities for optimisation but also presents challenges in terms of energy efficiency.

Nevertheless, the increase in data demands has significantly increased the power consumption of the next-generation networks, with an estimated  $75\%$ \ac{RAN} contributions \cite{green_ran_survey_2023}. Such an increase in power demands of the \ac{RAN} infrastructure has prompted significant research efforts in both academia and industry to develop energy-efficient solutions \cite{Lance_EE_ORAN_survey_2024, Imran_EE_survey_2023, ORAN_Use_Cases_Detailed_Specification_v11.00}. To address these challenges, the O-RAN Alliance has proposed several energy-saving techniques, including dynamic radio resource management, intelligent sleep modes, radio chain switch-off and AI-driven power optimization \cite{ORAN_EE_TR_v2.00}. Moreover, several researchers proposed algorithms to reduce the \ac{RAN} power consumption by placing \acp{RU} into sleep mode as in \cite{Lance_viavi_wincom_2024, Letter_EE_2016} or machine learning based algorithms \cite{Qiao_EE_camad_2024, korean_EE_RL_TWC_2022}. These methods aim to reduce the energy footprint of O-RAN networks by dynamically adapting power consumption to traffic demands while maintaining Quality of Service (QoS).


Considering these recommendations, several models have been proposed in the literature to profile \ac{RAN} power consumption, such as those presented in \cite{VIAVI_model_RAN} and \cite{Earth_model_RAN}. However, these models are designed for traditional \ac{RAN} architectures rather than O-RAN specific deployments. In the context of Open \ac{RAN} power consumption measurements, only a limited number of studies have conducted real-world evaluations. For instance, in \cite{POET_vtc_2024}, the power consumption of cloudified and virtualised network functions was assessed using various measurement techniques. Similarly, in \cite{EARNEST_2024}, the authors focused on CPU-level power consumption in virtualised RAN environments. In contrast, the study in \cite{EE_RU_access_2020} conducted power consumption measurements for \ac{RU} and \ac{DU} in an Open \ac{RAN}-based LTE system; however, it was limited to the uplink case and did not specify the functional split type.

This paper presents an experimental study on power consumption in O-RAN functional splits, specifically Split 8 and Split 7.2b, under varying \ac{PRB} utilisations. The Split 8 testbed is implemented using a \ac{SDR} USRP as the \ac{RU}, along with an srsRAN-based \cite{srsran} \ac{DU} and \ac{CU}. In contrast, the Split 7.2b testbed utilises commercial \acp{RU} from Benetel \cite{benetel}, designed for both indoor and outdoor deployments, providing a real-world evaluation of power consumption trends. To the best of the authors’ knowledge, this is the first study to conduct detailed power measurements for O-RAN functional splits and for both uplink and downlink, offering empirical insights into the energy efficiency of disaggregated RAN architectures. Furthermore, the measurements obtained in this study provide foundations for developing accurate digital twins of O-RAN deployments. 
By employing these measurements, a digital twin model can provide a better replica to real-world scenarios, enabling a more effective power optimisation strategies for energy efficient O-RAN management.



\begin{figure*}[!t]
    \centering
    \subfigure[Split 8 physical deployment]{%
        \includegraphics[clip, trim=0.0cm 2.5cm 0.0cm 2.0cm, width=1.5\columnwidth]{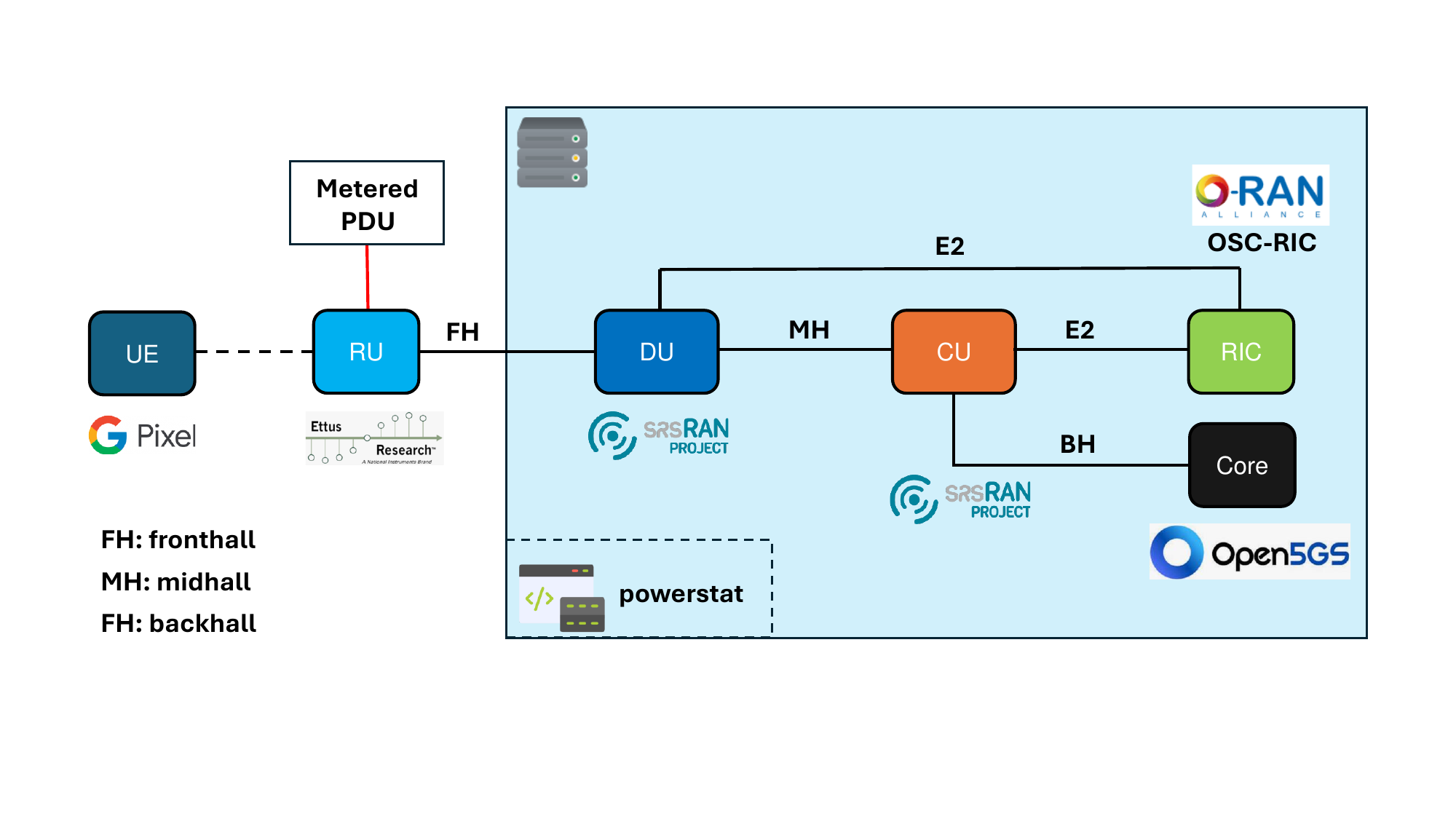}
        \label{fig:split_8_arch}
    }
    \hfill
    \subfigure[Split 7.2b physical deployment]{%
        \includegraphics[clip, trim=0.0cm 2.5cm 0.0cm 1cm, width=1.5\columnwidth]{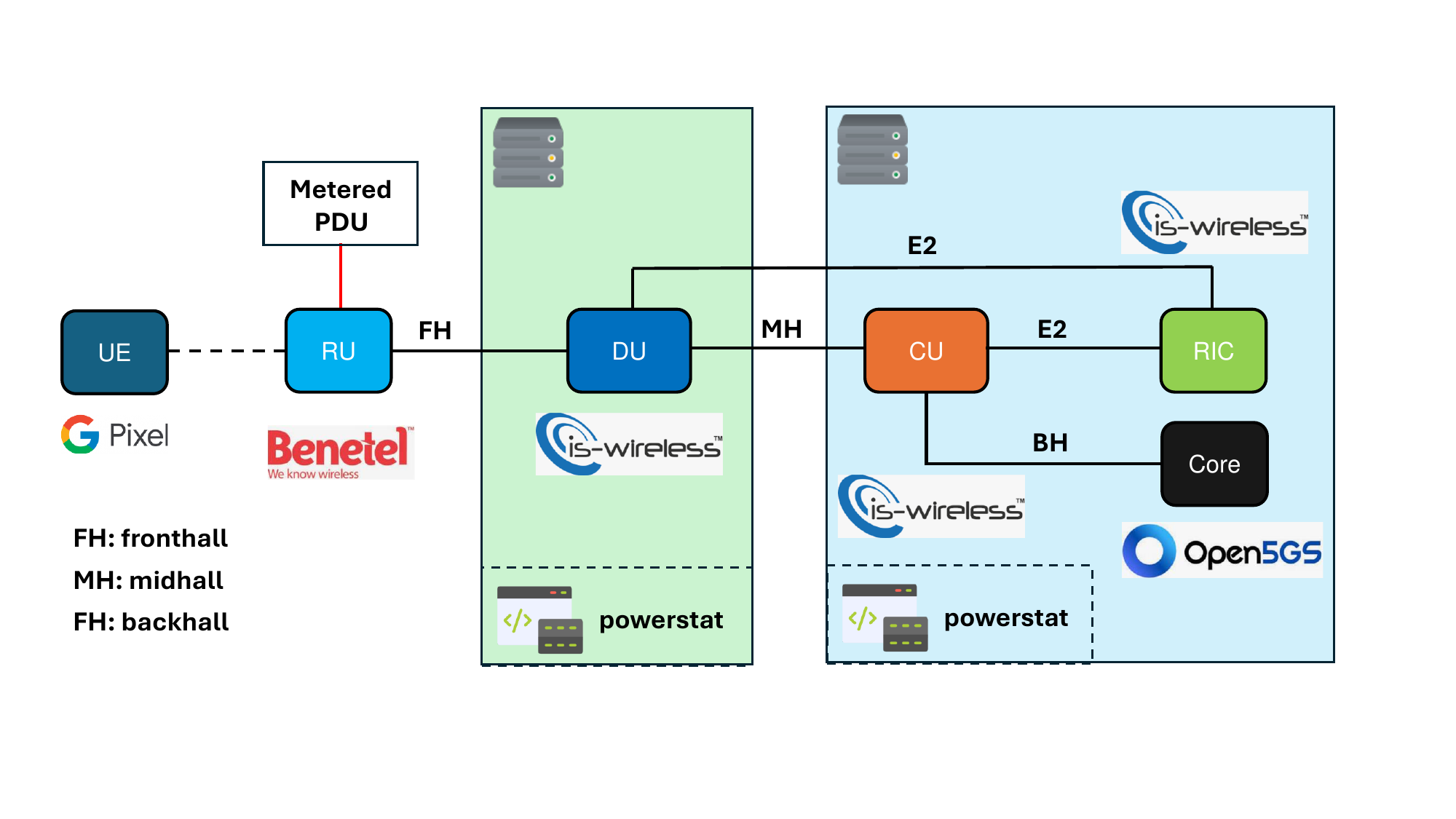}
        \label{fig:split_7_arch}
    }
    \caption{The physical deployments for different functional splits: (a) Split 8 and (b) Split 7.2b.}
    \label{fig:splits_arch}
    \vspace{-2mm}
\end{figure*}

\section{O-RAN functional splits}
\label{sec:functional_splits}
Open \ac{RAN} introduces flexibility and interoperability in mobile networks through the disaggregation of traditional \ac{RAN}  into distinct components (i.e., \ac{RU}, \ac{DU} and \ac{CU}) and therefore allows multiple functional split options. These splits impact latency, power consumption, fronthaul requirements and network flexibility \cite{ORAN_splits_2020}. Hence, such a modular approach is essential for an optimised deployment strategy tailored to specific network requirements. Among the various options, Split 8 and Split 7.2 with its variants have gained significant attention due to their distinct characteristics and deployment trade-offs \cite{OpenRAN_Splits_2025}.

Split 8 is shown in Fig. \ref{fig:split_8}, represents a separation between the \ac{PHY} and the \ac{RF} processing to maximise virtualisation gains \cite{RCRWireless_FunctionalSplits_2021}. The \ac{RU} handles only the \ac{RF} signal transmission and reception. While the \ac{DU} performs the baseband processing and the \ac{CU} is responsible for handling higher-layer protocols . This type of deployment enables use cases that require non-standard PHY signal processing such as the cell-free MIMO networks \cite{cellfreemimo_yi_york_2025}. 

Split 7.2b on the other hand, is illustrated in \ref{fig:split_7}, represents lower \ac{PHY} layer division, specifically separating the Low-\ac{PHY} and High-\ac{PHY} functions. The \ac{RU} handles the Low-\ac{PHY} functions, such as IFFT / FFT and beamforming tasks. While the \ac{DU} is responsible for High-\ac{PHY} functions, including coding/decoding and modulation/demodulation processes. While the \ac{CU} is responsible for the higher-layer protocol handling. Such type of deployment is suitable for use cases such as massive MIMO and Ultra-Reliable Low-Latency Communication (URLLC).

\section{Measurement testbeds}
In this section, we describe the experimental setup used to measure the power consumption of the O-RAN. The architecture of these testbeds is depicted in Fig. \ref{fig:splits_arch}. The server hosting the Split 8 system is a Dell PowerEdge R7515 (AMD EPYC 7662 2 GHz, 64 cores). The Split 7.2b system is hosted on two identical PowerEdge R760XA servers (Intel Xeon Gold 5420+ 2 GHz, 56 cores) with one server hosting the DU and the other hosting the other components.  The power measurements for the \acp{RU}, were conducted by querying a metered \ac{PDU}. 
While the power measurements for the bare metal servers with a docker image of \ac{CU}, \ac{DU}, \ac{RIC} and Core network utilising powerstat which is an open-source energy monitoring functionality for machines  (based on intel RAPL) and estimate the power consumed based on a power consumption model \cite{powerstat}. In both of the measurements campaigns, the RIC was deployed but without  any active xApps.

\subsection{Split 8}
The Split 8 O-RAN testbed is configured to operate with a 40 MHz bandwidth in the n78 band. The carrier frequency was set to 3.7 GHz. The testbed was designed to evaluate power consumption and performance across different O-RAN components, including the \ac{RU}, \ac{DU} and \ac{DU}, as illustrated in Fig. \ref{fig:split_8_arch}, which depicts the various components:

\begin{itemize}
\item \ac{RU}: The testbed employed a USRP, responsible for RF transmission and reception. The RU connects to the DU via a fibre fronthaul (FH).  

\item \ac{DU}: The srsRAN DU was deployed to manage low-MAC and PHY processing, interfacing with the RU via FH.  

\item \ac{CU}: The srsRAN CU was responsible for higher-layer RRC and PDCP processing, connected via Midhaul (MH) to the DU and Backhaul (BH) to the Core Network. The CU also interfaces with the RIC via the E2 interface.  

\item \ac{RIC}: The FlexRIC platform was deployed to enable policy control, energy efficiency optimisation and real-time monitoring. It communicates with the CU/DU over E2 interfaces.  

\item Core Network: The testbed employed Open5GS, handling authentication, session management and mobility control.  
\end{itemize}

The system was configured with a 6DS3U TDD pattern, allocating six slots for downlink (DL), three for uplink (UL) and one special slot for guard period or switching. The maximum Modulation and Coding Scheme (MCS) was configured to MCS 28 (64-QAM) for both uplink and downlink. A 2×2 MIMO configuration was deployed, utilising two transmit antennas (TX) and two receive antennas (RX) to improve spectral efficiency.

The maximum throughput achieved was 187 Mbps in the downlink and 42 Mbps in the uplink with MCS 28. The power consumption measurements obtained from the O-RU, O-DU and O-CU provide valuable insights into how power scales under different PRB utilisation levels, contributing to a better understanding of Split 8 energy efficiency in real-world deployments.

The testbed configurations are summarised in Table \ref{tab:Split_8_testbed_deployment_main_features}.

\subsection{Split 7.2x}
The 7.2b split O-RAN testbed is configured to operate with a 40 MHz bandwidth in the n78 and n77 bands. The carrier frequency was set to 3747.8  MHz (n78) for indoor testing and 3947.85 MHz (n77) for outdoor deployment. The testbed was designed to evaluate power consumption and performance across different O-RAN components, including the \ac{RU}, \ac{DU} and \ac{CU} as illustrated in Fig. \ref{fig:split_7_arch} which depicts the various components:
\begin{itemize}
    \item \ac{UE}: A Google Pixel 5G device was used.

    \item \acp{RU}: A Benetel RAN550 and a RAN650 were employed to handle the RF transmission and reception for indoor and outdoor setups, receptively. The RU connects to the DU via a fibre FH.

    \item \ac{DU}: The IS-Wireless DU was deployed to manage low-level MAC and PHY processing, interfacing with the RU over FH.

    \item \ac{CU}: The IS-Wireless CU was responsible for higher-layer RRC and PDCP processing, connected via MH to the DU and BH to the Core Network. The CU also interfaces with the RIC via the E2 interface.

    \item \ac{RIC}: The IS-Wireless RIC was deployed to enable policy control, energy efficiency optimisation and real-time monitoring. It communicates with the CU/DU over E2 interfaces.

    \item Core Network: The testbed employed Open5GS, handling authentication, session management and mobility control.

\end{itemize}

The system was configured with a 7DS2U TDD pattern. The maximum MCS was configured to 64-QAM for downlink and uplink. A 2$\times$2 MIMO configuration was deployed, utilising two TX antennas and two RX antennas. The testbed configurations are summarised in Table \ref{tab:Split_7_testbed_deployment_main_features}.

\begin{table}[!t]
\caption{Split 8 testbed deployment main features}
\label{tab:Split_8_testbed_deployment_main_features}
\resizebox{\columnwidth}{!}{%
\begin{tabular}{|c|c|}
\hline
\textbf{Feature}        & \textbf{Description}    \\ \hline
Frequency band          & n78 (FR1, TDD)          \\ \hline
Carrier frequency       & 3.7 GHz                 \\ \hline
Gain                    & 30 dB                     \\ \hline
Bandwidth               & 40 MHz                  \\ \hline
Subcarrier spacing      & 30 KHz                  \\ \hline
TDD config              & 6DS3U                   \\ \hline
Number of antennas used & 2 TX, 2 RX              \\ \hline
MIMO config             & 2 layers DL, 1 layer UL \\ \hline
Maximum MCS& 28 (i.e., 64-QAM)       \\ \hline
Max throughput uplink   & 42 Mbps                 \\ \hline
Max throughput downlink & 187 Mbps                \\ \hline
\end{tabular}%
}
\end{table}

\begin{table}[!t]
\caption{Split 7.2b testbed deployment main features}
\label{tab:Split_7_testbed_deployment_main_features}
\resizebox{\columnwidth}{!}{%
\begin{tabular}{|c|c|}
\hline
\textbf{Feature}        & \textbf{Description}                           \\ \hline
Frequency band          & n78/n77 (FR1, TDD)\\ \hline
Carrier frequency       & \makecell{3747.8 MHz for indoor\\ 3947.85 MHz for outdoor} \\ \hline
Gain                    & 24 dB                                          \\ \hline
Bandwidth               & 40 MHz                                         \\ \hline
Subcarrier spacing      & 30 KHz                                         \\ \hline
TDD config              & 7DS2U                                          \\ \hline
Number of antennas used & 2 TX, 2 RX                                     \\ \hline
MIMO config             & 2 layers DL, 1 layer UL                        \\ \hline
Maximum MCS& 28 (i.e., 64-QAM)                              \\ \hline
Max throughput uplink   & 26 Mbps                                        \\ \hline
Max throughput downlink & 177 Mbps                                       \\ \hline
\end{tabular}%
}
\end{table}

\section{Experimental measurements}
\label{exp_results}
In this section, we present the measurement results from the O-RAN testbeds. Subsection \ref{split_8_measurment} introduces the measurements for split 8 with USRP serving as the \ac{RU} and srsRAN CU and DU. Subsection \ref{split_7_measurment} introduces the measurements for split 7.2 (split 7.2b specifically which is a variation of split 7.2) including two commercial \acp{RU} for indoor and outdoor with commercial CU and DU. We  conducted the experiments inside the Institute for Safe Autonomy at the University of York. The lab is a 100
m$^2$ indoor lab and the \ac{UE} device was placed within 3 m of the \ac{RU} for better signal quality.

\subsection{Split 8 measurements}
\label{split_8_measurment}
Table \ref{tab:split_8_RUs} shows the power consumption of the uplink and downlink transmissions, respectively across different \ac{PRB} utilisation levels.  The uplink power consumption remains relatively stable, ranging between 43.1 W and 44.2 W, with minimal variations in the standard deviation (STD). Notably, there is no clear linear correlation between \ac{PRB} utilisation and uplink power consumption, while the downlink shows a marginal increase. Such results are expected given that minimal signal processing occurs in the split 8 \ac{RU} as defined by the O-RAN Alliance.
Please note that 0$\%$ \ac{PRB} utilisation refers to zero throughput (i.e., zero \acp{UE} attached). Therefore, we have the same uplink and downlink values.

Next, Table \ref{tab:split_8_CU_DU} presents the combined power consumption of the CU and DU for split 8. Idle refers to having the server running alone without deploying the \ac{CU} and \ac{DU} (i.e., only Linux background process). The 0$\%$ \ac{PRB} utilisation refers to the case that no \acp{UE} are attached to the network. In Table \ref{tab:split_8_CU_DU}, at idle, the system consumes 58.34 W and the energy consumption was doubled when the \ac{CU} and \ac{DU} were switched on but not serving any \acp{UE}, indicating a significant base power requirement to have the network operating. Furthermore, as the \ac{PRB} utilisation increases, both uplink and downlink power consumption increases steadily. With downlink requiring more power compared to uplink with an almost 15 W difference when in full load (i.e., \ac{PRB} utilisation 100$\%$). 


\subsection{Split 7.2b measurements}
\label{split_7_measurment}
This subsection will introduce split 7.2b measurement results for both indoor and outdoor \acp{RU}, in addition to the \ac{CU} and \ac{DU}. Table \ref{tab:split_7_RUs} shows the measurements for the Benetel RAN550, a commercial indoor \ac{RU} and Benetel RAN650 a commercial outdoor \ac{RU}. 

For the Benetel RAN550, the uplink power remains nearly constant at 29.0 W, while the downlink increases slightly to 30.1 W at full \ac{PRB} utlisation. Similarly, for the Benetel RAN650, the uplink power remains at 46.0 W, while the downlink rises marginally to 46.2 W at 100$\%$ PRB utilisation. Compared to Split 8, the power consumption in Split 7.2b is more stable, with minimal variations across different loads, which is expected due to using \ac{SDR} as \ac{RU} for split 8 configuration. 

Tables \ref{tab:split_7_DU} and \ref{tab:split_7_CU} present the power consumption of the \ac{CU} and \ac{DU} in split 7.2 across different PRB utilisation levels. In the DU (Table \ref{tab:split_7_DU}), the power consumption starts at 187.2 W in idle mode and increases gradually with PRB utilisation, reaching 193.15 W (uplink) and 194.21 W (downlink) at full load. The power variance remains low, indicating stable performance. Similarly, in the \ac{CU} (Table \ref{tab:split_7_CU}), power consumption starts at 189.6 W when idle and increases slightly to 190.97 W (uplink) and 192.67 W (downlink) at maximum \ac{PRB} utilisation. The \ac{CU}’s power consumption is relatively stable, with minimal variation across different PRB levels. It should be noted that the power increase for both \ac{CU} and \ac{DU} is minimal, with variations remaining within a range of 6 W across all PRB utilisation levels.


\begin{table}[!t]
\caption{Power consumption of a split 8 RU  under varying PRB utilisation}
\label{tab:split_8_RUs}
\resizebox{\columnwidth}{!}{%
\begin{tabular}{|c|cc|cc|}
\hline
\multirow{2}{*}{\textbf{\begin{tabular}[c]{@{}c@{}}PRB utilisation\\ in percentages\end{tabular}}} & \multicolumn{2}{c|}{\textbf{Uplink in Watts}}     & \multicolumn{2}{c|}{\textbf{Downlink in Watts}}   \\ \cline{2-5} 
                                                                                                   & \multicolumn{1}{c|}{\textbf{Mean}} & \textbf{STD} & \multicolumn{1}{c|}{\textbf{Mean}} & \textbf{STD} \\ \hline
0                                                                                                  & \multicolumn{1}{c|}{43.9}          & 2.4          & \multicolumn{1}{c|}{43.9}          & 2.4          \\ \hline
25                                                                                                 & \multicolumn{1}{c|}{43.3}          & 3.0          & \multicolumn{1}{c|}{44.9}          & 0.6          \\ \hline
50                                                                                                 & \multicolumn{1}{c|}{44.2}          & 3.4          & \multicolumn{1}{c|}{44.8}          & 1.1          \\ \hline
75                                                                                                 & \multicolumn{1}{c|}{43.1}          & 2.7          & \multicolumn{1}{c|}{44.8}          & 1.2          \\ \hline
100                                                                                                & \multicolumn{1}{c|}{44.2}          & 2.4          & \multicolumn{1}{c|}{45.0}          & 1.5          \\ \hline
\end{tabular}%
}
\end{table}

\begin{table}[!t]
\centering
\caption{Power consumption of split 8 CU and DU across different PRB utilisation levels.}
\label{tab:split_8_CU_DU}
\resizebox{\columnwidth}{!}{%
\begin{tabular}{|c|cc|cc|}
\hline
\multirow{2}{*}{\textbf{\begin{tabular}[c]{@{}c@{}}PRB utilisation\\ in percentages\end{tabular}}} & \multicolumn{2}{c|}{\textbf{Uplink in Watts}}     & \multicolumn{2}{c|}{\textbf{Downlink in Watts}}   \\ \cline{2-5} 
                                                                                                   & \multicolumn{1}{c|}{\textbf{Mean}} & \textbf{STD} & \multicolumn{1}{c|}{\textbf{Mean}} & \textbf{STD} \\ \hline
Idle                                                                                               & \multicolumn{1}{c|}{58.34}         & 4.7          & \multicolumn{1}{c|}{58.34}         & 4.7          \\ \hline
0                                                                                                  & \multicolumn{1}{c|}{119.51}        & 4.02         & \multicolumn{1}{c|}{119.51}        & 4.02         \\ \hline
25                                                                                                 & \multicolumn{1}{c|}{120.65}        & 3.8          & \multicolumn{1}{c|}{123.48}        & 4.2          \\ \hline
50                                                                                                 & \multicolumn{1}{c|}{121.21}        & 3.93         & \multicolumn{1}{c|}{125.37}        & 4.14         \\ \hline
75                                                                                                 & \multicolumn{1}{c|}{124.7}         & 3.62         & \multicolumn{1}{c|}{128.62}        & 4.83         \\ \hline
100                                                                                                & \multicolumn{1}{c|}{125.19}        & 4.05         & \multicolumn{1}{c|}{141.59}        & 4.57         \\ \hline
\end{tabular}%
}
\end{table}

\begin{table}[]
\caption{Power consumption for split 7.2b RUs under varying PRB utilisation}
\label{tab:split_7_RUs}
\resizebox{\columnwidth}{!}{%
\begin{tabular}{|c|cc|cc|}
\hline
\multirow{2}{*}{\textbf{\begin{tabular}[c]{@{}c@{}}PRB utilisation\\ in percentages\end{tabular}}} & \multicolumn{2}{c|}{\textbf{\begin{tabular}[c]{@{}c@{}}Benetel 550 power \\ consumption in Watts\end{tabular}}} & \multicolumn{2}{c|}{\textbf{\begin{tabular}[c]{@{}c@{}}Benetel 650 power\\ consumption in Watts\end{tabular}}} \\ \cline{2-5} 
                                                                                                   & \multicolumn{1}{c|}{\textbf{Uplink}}                             & \textbf{Downlink}                            & \multicolumn{1}{c|}{\textbf{Uplink}}                            & \textbf{Downlink}                            \\ \hline
0                                                                                                  & \multicolumn{1}{c|}{28.3}                                        & 28.3                                         & \multicolumn{1}{c|}{44.6}                                       & 44.6                                         \\ \hline
25                                                                                                 & \multicolumn{1}{c|}{29.0}                                        & 29.0                                         & \multicolumn{1}{c|}{46.0}                                       & 46.1                                         \\ \hline
50                                                                                                 & \multicolumn{1}{c|}{29.0}                                        & 30.0                                         & \multicolumn{1}{c|}{46.0}                                       & 46.1                                         \\ \hline
75                                                                                                 & \multicolumn{1}{c|}{29.0}                                        & 30.1                                         & \multicolumn{1}{c|}{46.0}                                       & 46.1                                         \\ \hline
100                                                                                                & \multicolumn{1}{c|}{29.0}                                        & 30.1                                         & \multicolumn{1}{c|}{46.0}                                       & 46.2                                         \\ \hline
\end{tabular}%
}
\end{table}

\begin{table}[!t]
\caption{Power consumption of split 7.2b DU under varying PRB utilisation}
\label{tab:split_7_DU}
\resizebox{\columnwidth}{!}{%
\begin{tabular}{|c|cc|cc|}
\hline
\multirow{2}{*}{\textbf{\begin{tabular}[c]{@{}c@{}}PRB utilisation\\ in percentages\end{tabular}}} & \multicolumn{2}{c|}{\textbf{Uplink in Watts}}     & \multicolumn{2}{c|}{\textbf{Downlink in Watts}}   \\ \cline{2-5} 
                                                                                                   & \multicolumn{1}{c|}{\textbf{Mean}} & \textbf{STD} & \multicolumn{1}{c|}{\textbf{Mean}} & \textbf{STD} \\ \hline
Idle                                                                                               & \multicolumn{1}{c|}{187.2}         & 0.13         & \multicolumn{1}{c|}{187.2}         & 0.13         \\ \hline
0                                                                                                  & \multicolumn{1}{c|}{191.8}        & 0.42         & \multicolumn{1}{c|}{191.8}        & 0.42         \\ \hline
25                                                                                                 & \multicolumn{1}{c|}{191.9}        & 0.47         & \multicolumn{1}{c|}{192.6}        & 0.44         \\ \hline
50                                                                                                 & \multicolumn{1}{c|}{192.6}        & 0.48         & \multicolumn{1}{c|}{192.8}        & 0.43         \\ \hline
75                                                                                                 & \multicolumn{1}{c|}{193.1}        & 0.46         & \multicolumn{1}{c|}{193.7}        & 0.42         \\ \hline
100                                                                                                & \multicolumn{1}{c|}{193.2}        & 0.62         & \multicolumn{1}{c|}{194.2}        & 0.39         \\ \hline
\end{tabular}%
}
\end{table}

\begin{table}[!t]
\caption{Power consumption of split 7.2b CU under varying PRB utilisation}
\label{tab:split_7_CU}
\resizebox{\columnwidth}{!}{%
\begin{tabular}{|c|cc|cc|}
\hline
\multirow{2}{*}{\textbf{\begin{tabular}[c]{@{}c@{}}PRB utilisation\\ in percentages\end{tabular}}} & \multicolumn{2}{c|}{\textbf{Uplink in Watts}}     & \multicolumn{2}{c|}{\textbf{Downlink in Watts}}   \\ \cline{2-5} 
                                                                                                   & \multicolumn{1}{c|}{\textbf{Mean}} & \textbf{STD} & \multicolumn{1}{c|}{\textbf{Mean}} & \textbf{STD} \\ \hline
Idle                                                                                               & \multicolumn{1}{c|}{189.6}         & 0.19         & \multicolumn{1}{c|}{189.6}         & 0.19         \\ \hline
0                                                                                                  & \multicolumn{1}{c|}{189.8}        & 0.29         & \multicolumn{1}{c|}{189.8}        & 0.29         \\ \hline
25                                                                                                 & \multicolumn{1}{c|}{189.8}         & 0.24         & \multicolumn{1}{c|}{190.6}        & 0.31         \\ \hline
50                                                                                                 & \multicolumn{1}{c|}{189.9}        & 0.26         & \multicolumn{1}{c|}{191.5}        & 0.28         \\ \hline
75                                                                                                 & \multicolumn{1}{c|}{190.1}        & 0.27         & \multicolumn{1}{c|}{192.0}        & 0.36         \\ \hline
100                                                                                                & \multicolumn{1}{c|}{191.0}        & 0.3          & \multicolumn{1}{c|}{192.7}        & 0.36         \\ \hline
\end{tabular}%
}
\end{table}

\section{Modelling the power consumption in Open RAN}
\begin{figure}[!t]
	\centering
	\includegraphics[clip, trim=0.0cm 0.cm 0.0cm 0cm, width=1\columnwidth]{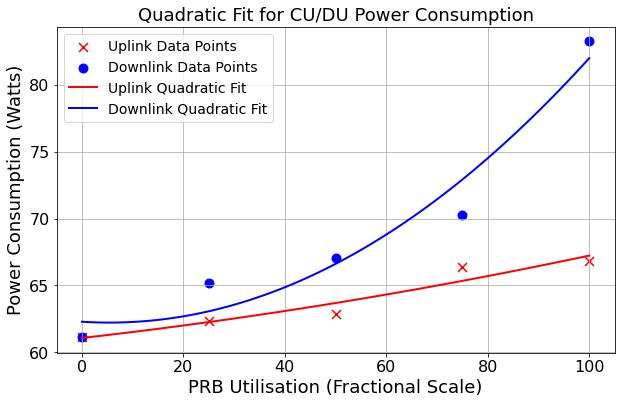}
	\caption{Comparison between measured and quadratic fit power model for split 8 DU and CU.}
	\label{fig:quad_fit}
\end{figure}

After obtaining the measured power consumption data, we can now model it effectively. The power consumption results for the split 7.2b \ac{RAN} and split's \ac{RU} were consistent, as presented in Section \ref{exp_results}. Therefore, we focus on split 8 combined \ac{DU} and \ac{CU} power consumptions. The quadratic polynomial fits are given as:

\begin{equation}
P^{DU\&CU}_{UL} = 1.86l^{2} + 4.3l +61.06
\end{equation}

\begin{equation}
P^{DU\&CU}_{DL} = 22.12l^{2} - 2.4l + 62.28
\end{equation}
where $l$ represents the \ac{PRB} utilisation percentage. The terms \( P^{DU\&CU}_{UL} \) and \( P^{DU\&CU}_{DL} \) denote the power consumption for the combined \ac{DU} and \ac{CU} in the uplink and downlink, respectively. Fig. \ref{fig:quad_fit} illustrates the measured power consumption alongside the quadratic model fit. As observed, the downlink exhibits higher power consumption than the uplink, which aligns with expectations due to increased transmission demands in the downlink in terms of throughput.

\section{Conclusions}
In this work, a comprehensive measurements and analysis of power consumption in O-RAN functional splits, focusing on Split 8 and Split 7.2b. The empirical results showed that the power consumption in O-RAN components, including RUs, DUs and CUs, does not scale proportionally with PRB utilisation and a significant portion of energy consumption remains constant regardless of network load. This highlights the need for energy-efficient strategies in O-RAN deployments to focus on switching on/off radio chain or RAN components. In addition, the results showed that downlink power consumption is consistently higher than uplink, which aligns with the increased processing and transmission demands in throughput for downlink. 

The empirical measurements obtained in this study not only aid in developing energy-saving solutions but also contribute to the creation of a digital twin for the network. By leveraging these insights, advanced power-saving mechanisms such as intelligent resource management, dynamic power adaptation, AI-driven optimization techniques, and digital twin-based energy modelling can be effectively implemented. Digital twin technology, in particular, plays a crucial role in real-time network monitoring, power forecasting, and proactive energy management, enabling operators to simulate different configurations and dynamically optimize power consumption. These solutions collectively help mitigate the energy footprint of O-RAN deployments while ensuring network performance and service quality.

\section*{Acknowledgment}
This work was supported in part by the Engineering and
Physical Sciences Research Council United Kingdom (EPSRC), Impact Acceleration Accounts (IAA) (Green Secure and
Privacy Aware Wireless Networks for Sustainable Future Connected and Autonomous Systems) under Grant EP/X525856/1
and Department of Science, Innovation and Technology,
United Kingdom, under Grants Yorkshire Open-RAN (YORAN) TS/X013758/1 and RIC Enabled (CF-)mMIMO for
HDD (REACH) TS/Y008952/1.

\begin{acronym} 
\acro{3GPP}{3rd Generation Partnership Project}
\acro{5G}{Fifth Generation}
\acro{AI}{Artificial Intelligence}
\acro{ANN}{Artificial Neural Network}
\acro{API}{Application Programming Interface}
\acro{BBU}{Base Band Unit}
\acro{BER}{Bit Error Rate}
\acro{BS}{Base Station}
\acro{BW}{bandwidth}
\acro{C-RAN}{Cloud Radio Access Networks}
\acro{CDMA}{Code Division Multiple Access}
\acro{CNN}{Conventional Neural Network}
\acro{CoMP}{Coordinated Multipoint}
\acro{COTS}{Commercial off-the-shelf}
\acro{CP}{Control Plane}
\acro{CR}{Cognitive Radio}
\acro{CU}{Central Unit}
\acro{DAC}{Digital-to-Analog Converter}
\acro{D2D}{Device-to-Device}
\acro{DAS}{Distributed Antenna Systems}
\acro{DC}{Duty Cycle}
\acro{DDQN}{Double Deep Q-Learning Network}
\acro{DL}{Deep Learning}
\acro{DNN}{Deep Neural Network}
\acro{DQN}{Deep Q-Learning Network}
\acro{DRL}{Deep Reinforcement Learning}
\acro{D-RAN}{Distributed RAN}
\acro{DSA}{Dynamic Spectrum Access}
\acro{DU}{Distributed Unit}
\acro{EE}{Energy Efficiency}
\acro{eMMB}{enhanced Mobile Broadband}
\acro{ES}{Energy Saving}
\acro{FDD}{Frequency Division Duplex}
\acro{FEC}{Forward Error Correction}
\acro{FFT}{Fast Fourier Transform}
\acro{FFR}{Fractional Frequency Reuse}
\acro{FPGAs}{Field  Programmable  Gate  Arrays}
\acro{FSPL}{Free Space Path Loss}
\acro{GA}{Genetic Algorithms}
\acro{GNN}{Graph Neural Network}
\acro{HARQ}{Hybrid-Automatic Repeat Request}
\acro{HetNet}{Heterogeneous Network}
\acro{HO}{Handover}
\acro{HTTP}{Hypertext Transfer Protocol}
\acro{IC}{Immersive Communication}
\acro{ICA}{Independent Component Analysis}
\acro{ILP}{Integer Linear Programming}
\acro{IoT}{Internet of Things}
\acro{KNN}{k-Nearest Neighbors}
\acro{KPM}{Key Performance Measurements}
\acro{KPI}{Key Performance Indicators}
\acro{LAN}{Local Area Network}
\acro{LOS}{Line of Sight}
\acro{LTE}{Long Term Evolution}
\acro{LTE-A}{Long Term Evolution Advanced}
\acro{LSTM}{Long Short-term Memory}
\acro{MAC}{Medium Access Control}
\acro{MDP}{Markov Decision Process}
\acro{ML}{Machine Learning}
\acro{MLP}{Multiple-layer Perceptron}
\acro{mMTC}{massive Machine-Type Communication}
\acro{MIMO}{Multiple-Input Multiple-Output}
\acro{m-MIMO}{massive Multiple-Input Multiple-Output}
\acro{mmWave}{millimeter Wave}
\acro{Near-RT}{Near-Real Time}
\acro{Near-RT RIC}{Near Real Time RIC}
\acro{NFV}{Network Function Virtualization}
\acro{NIB}{Network Information Base}
\acro{NLoS}{Non-Line of Sight}
\acro{NN}{Neural Network}
\acro{NR}{New Radio}
\acro{OFDM}{Orthogonal Frequency Division Multiplexing}
\acro{OFDMA}{Orthogonal Frequency-Division Multiple Access}
\acro{O-RAN}{Open Radio Access Network }
\acro{Open RAN}{Open Radio Access Network}
\acro{OPEX}{Operational Expenditure}
\acro{OSC}{O-RAN Software Community}
\acro{OSA}{Opportunistic Spectrum Access}
\acro{PA}{Power Amplifier}
\acro{PAM}{Pulse Amplitude Modulation}
\acro{PAPR}{Peak-to-Average Power Ratio}
\acro{PCA}{Principal Component Analysis}
\acro{PDCP}{Packet Data Convergence Protocol}
\acro{PDU}{Power Distribution Unit}
\acro{PG}{Policy Gradient}
\acro{PHY}{Physical layer}
\acro{PRB}{Physical resource block}
\acro{PSO}{Particle Swarm Optimization}
\acro{PU}{Primary User}
\acro{QL}{Q-Learning}
\acro{QAM}{Quadrature Amplitude Modulation}
\acro{QoE}{Quality of Experience}
\acro{QoS}{Quality of Service}
\acro{QPSK}{Quadrature Phase Shift Keying}
\acro{RAN}{Radio Access Network}
\acro{RAT}{Radio Access Technology}
\acro{RC}{Radio Card}
\acro{RF}{Radio Frequency}
\acro{RIC}{RAN Intelligent Controller}
\acro{RLC}{Radio Link Control}
\acro{RL}{Reinforcement Learning}
\acro{RMSE}{Root Mean Squared Error}
\acro{RN}{Remote Node}
\acro{RRH}{Remote Radio Head}
\acro{RRM}{Radio Resources Management}
\acro{RRC}{Radio Resource Control}
\acro{RRU}{Remote Radio Unit}
\acro{RSS}{Received Signal Strength}
\acro{RSRP}{Reference Signals Received Power}
\acro{RT}{Real Time}
\acro{RU}{Radio Unit}
\acro{SCA}{Successive Convex Approximation}
\acro{SCBS}{Small Cell Base Station}
\acro{SDL}{Shared Data Layer}
\acro{SDN}{Software Defined Network}
\acro{SDR}{Software Defined Radio}
\acro{SDS}{Software Defined Security}
\acro{SDAP}{Service Data Adaptation Protocol}
\acro{SHAP}{SHapley Additive exPlanations}
\acro{SINR}{Signal-to-Interference-plus-Noise Ratio}
\acro{SLA}{Service Level Agreement}
\acro{SMO}{Service Management and Orchestration}
\acro{SNR}{Signal-to-Noise Ratio}
\acro{SON}{Self-organised Network}
\acro{SU}{Secondary User}
\acro{SU-MIMO}{Single-User MIMO}
\acro{SVM}{Support Vector Machine}
\acro{TD}{Temporal Defence}
\acro{TDD}{Time Division Duplex}
\acro{TD-LTE}{Time Division LTE}
\acro{TDMA}{Time Division Multiple Access}
\acro{TS}{Traffic Steering}
\acro{UDN}{Ultra-Dense Network}
\acro{UE}{User Equipment}
\acro{UMa}{Urban Macrocell path loss}
\acro{UMi}{Urban Microcell path loss}
\acro{URLLC}{ultra-reliable low-latency communication}
\acro{USRP}{Universal Software Radio Platform}
\acro{VFN}{Network Function Virtualization}
\acro{v-RAN}{Virtual RAN}
\acro{XAI}{eXplainable Artificial Intelligent}
\end{acronym}

\footnotesize
\bibliographystyle{IEEEtran}
\bibliography{References}

\begin{thebibliography}{10}
\providecommand{\url}[1]{#1}
\csname url@samestyle\endcsname
\providecommand{\newblock}{\relax}
\providecommand{\bibinfo}[2]{#2}
\providecommand{\BIBentrySTDinterwordspacing}{\spaceskip=0pt\relax}
\providecommand{\BIBentryALTinterwordstretchfactor}{4}
\providecommand{\BIBentryALTinterwordspacing}{\spaceskip=\fontdimen2\font plus
\BIBentryALTinterwordstretchfactor\fontdimen3\font minus \fontdimen4\font\relax}
\providecommand{\BIBforeignlanguage}[2]{{%
\expandafter\ifx\csname l@#1\endcsname\relax
\typeout{** WARNING: IEEEtran.bst: No hyphenation pattern has been}%
\typeout{** loaded for the language `#1'. Using the pattern for}%
\typeout{** the default language instead.}%
\else
\language=\csname l@#1\endcsname
\fi
#2}}
\providecommand{\BIBdecl}{\relax}
\BIBdecl

\bibitem{IC_frontiers_survy_2023}
\BIBentryALTinterwordspacing
X.~S. Shen, J.~Gao, M.~Li, C.~Zhou, S.~Hu, M.~He, and W.~Zhuang, ``Toward immersive communications in {6G},'' \emph{Frontiers in Computer Science}, vol.~4, 2023. [Online]. Available: \url{https://www.frontiersin.org/articles/10.3389/fcomp.2022.1068478/full}
\BIBentrySTDinterwordspacing

\bibitem{Ericsson5G}
\BIBentryALTinterwordspacing
Ericsson, ``Energy consumption of 5g new radio,'' 2019, accessed: 2025-03-06. [Online]. Available: \url{https://www.ericsson.com/en/blog/2019/9/energy-consumption-5g-nr}
\BIBentrySTDinterwordspacing

\bibitem{ORAN_Architecture_2022}
\BIBentryALTinterwordspacing
{O-RAN Alliance}, ``{O-RAN Architecture Description v03.00},'' O-RAN Technical Specification, 2022, accessed: 2025-03-06. [Online]. Available: \url{https://www.o-ran.org/specifications}
\BIBentrySTDinterwordspacing

\bibitem{green_ran_survey_2023}
\BIBentryALTinterwordspacing
L.~M.~P. Larsen, H.~Christiansen, S.~Ruepp, and M.~Berger, ``Toward greener {5G} and beyond radio access networks—a survey,'' \emph{IEEE Open Journal of the Communications Society}, vol.~4, pp. 569--590, 2023. [Online]. Available: \url{https://ieeexplore.ieee.org/document/10076806/}
\BIBentrySTDinterwordspacing

\bibitem{Lance_EE_ORAN_survey_2024}
X.~Liang, Q.~Wang, A.~Al-Tahmeesschi, S.~B. Chetty, D.~Grace, and H.~Ahmadi, ``Energy consumption of machine learning enhanced open ran: A comprehensive review,'' \emph{IEEE Access}, vol.~12, pp. 81\,889--81\,910, 2024.

\bibitem{Imran_EE_survey_2023}
A.~I. Abubakar, O.~Onireti, Y.~Sambo, L.~Zhang, G.~K. Ragesh, and M.~Ali~Imran, ``Energy efficiency of open radio access network: A survey,'' in \emph{2023 IEEE 97th Vehicular Technology Conference (VTC2023-Spring)}, 2023, pp. 1--7.

\bibitem{ORAN_Use_Cases_Detailed_Specification_v11.00}
\BIBentryALTinterwordspacing
{O-RAN Alliance Work Group 1}, ``{O-RAN Use Cases Detailed Specification v11.00},'' O-RAN Alliance, Tech. Rep., 2023, accessed: 2025-03-07. [Online]. Available: \url{https://www.o-ran.org/specifications}
\BIBentrySTDinterwordspacing

\bibitem{ORAN_EE_TR_v2.00}
\BIBentryALTinterwordspacing
------, ``{O-RAN Network Energy Saving Use Cases Technical Report v2.00},'' O-RAN Alliance, Tech. Rep., 2023, accessed: 2025-03-07. [Online]. Available: \url{https://orandownloadsweb.azurewebsites.net/specifications}
\BIBentrySTDinterwordspacing

\bibitem{Lance_viavi_wincom_2024}
X.~Liang, A.~Al-Tahmeesschi, Q.~Wang, S.~Chetty, C.~Sun, and H.~Ahmadi, ``Enhancing energy efficiency in {O-RAN} through intelligent xapps deployment,'' in \emph{2024 11th International Conference on Wireless Networks and Mobile Communications (WINCOM)}, 2024, pp. 1--6.

\bibitem{Letter_EE_2016}
E.~Oh and K.~Son, ``A unified base station switching framework considering both uplink and downlink traffic,'' \emph{IEEE Wireless Communications Letters}, vol.~6, no.~1, pp. 30--33, 2017.

\bibitem{Qiao_EE_camad_2024}
\BIBentryALTinterwordspacing
Q.~Wang, S.~Chetty, A.~Al-Tahmeesschi, X.~Liang, Y.~Chu, and H.~Ahmadi, ``Energy saving in {6G} o-ran using dqn-based xapp,'' 2024. [Online]. Available: \url{https://arxiv.org/abs/2409.15098}
\BIBentrySTDinterwordspacing

\bibitem{korean_EE_RL_TWC_2022}
H.~Ju, S.~Kim, Y.~Kim, and B.~Shim, ``Energy-efficient ultra-dense network with deep reinforcement learning,'' \emph{IEEE Transactions on Wireless Communications}, vol.~21, no.~8, pp. 6539--6552, 2022.

\bibitem{VIAVI_model_RAN}
E.~Björnson, L.~Sanguinetti, J.~Hoydis, and M.~Debbah, ``Optimal design of energy-efficient multi-user {MIMO} systems: Is massive {MIMO} the answer?'' \emph{IEEE Transactions on Wireless Communications}, vol.~14, no.~6, pp. 3059--3075, 2015.

\bibitem{Earth_model_RAN}
G.~Auer, V.~Giannini, C.~Desset, I.~Godor, P.~Skillermark, M.~Olsson, M.~A. Imran, D.~Sabella, M.~J. Gonzalez, O.~Blume, and A.~Fehske, ``How much energy is needed to run a wireless network?'' \emph{IEEE Wireless Communications}, vol.~18, no.~5, pp. 40--49, 2011.

\bibitem{POET_vtc_2024}
N.~K. Shankaranarayanan, Z.~Li, I.~Seskar, P.~Maddala, S.~Puthenpura, A.~Stancu, and A.~Agarwal, ``{POET}: A platform for {O-RAN} energy efficiency testing,'' in \emph{2024 IEEE 100th Vehicular Technology Conference (VTC2024-Fall)}, 2024, pp. 1--5.

\bibitem{EARNEST_2024}
V.~Gudepu, B.~Chirumamilla, R.~R. Tella, A.~Bhattacharyya, S.~Agarwal, L.~Malakalapalli, C.~Centofanti, J.~Santos, and K.~Kondepu, ``{EARNEST}: Experimental analysis of {RAN} energy with open-source software tools,'' in \emph{2024 16th International Conference on COMmunication Systems $\&$ NETworkS (COMSNETS)}, 2024, pp. 1148--1153.

\bibitem{EE_RU_access_2020}
S.~Wu, Y.~Wang, and L.~Bai, ``Deep convolutional neural network assisted reinforcement learning based mobile network power saving,'' \emph{IEEE Access}, vol.~8, pp. 93\,671--93\,681, 2020.

\bibitem{srsran}
\BIBentryALTinterwordspacing
{Software Radio Systems}, ``{srsRAN},'' 2025, accessed: 2025-03-07. [Online]. Available: \url{https://www.srslte.com/5g}
\BIBentrySTDinterwordspacing

\bibitem{benetel}
\BIBentryALTinterwordspacing
{Benetel Ltd}, ``{Benetel},'' 2025, accessed: 2025-03-07. [Online]. Available: \url{https://benetel.com/}
\BIBentrySTDinterwordspacing

\bibitem{ORAN_splits_2020}
V.~Q. Rodriguez, F.~Guillemin, A.~Ferrieux, and L.~Thomas, ``Cloud-{RAN} functional split for an efficient fronthaul network,'' in \emph{2020 International Wireless Communications and Mobile Computing (IWCMC)}, 2020, pp. 245--250.

\bibitem{OpenRAN_Splits_2025}
\BIBentryALTinterwordspacing
{5G Technology World}, ``Open {RAN} functional splits, explained,'' 2025, accessed: 2025-03-07. [Online]. Available: \url{https://www.5gtechnologyworld.com/open-ran-functional-splits-explained/}
\BIBentrySTDinterwordspacing

\bibitem{RCRWireless_FunctionalSplits_2021}
\BIBentryALTinterwordspacing
{RCR Wireless News}, ``Exploring functional splits in {5G RAN}: Tradeoffs and use cases,'' 2021, accessed: 2025-03-07. [Online]. Available: \url{https://www.rcrwireless.com/20210317/5g/exploring-functional-splits-in-5g-ran-tradeoffs-and-use-cases-reader-forum}
\BIBentrySTDinterwordspacing

\bibitem{cellfreemimo_yi_york_2025}
\BIBentryALTinterwordspacing
Y.~Chu, M.~Rahmani, J.~Shackleton, D.~Grace, K.~Cumanan, H.~Ahmadi, and A.~Burr, ``Testbed development: An intelligent {O-RAN} based cell-free {MIMO} network,'' 2025. [Online]. Available: \url{https://arxiv.org/abs/2502.08529}
\BIBentrySTDinterwordspacing

\bibitem{powerstat}
\BIBentryALTinterwordspacing
C.~I. King, ``Powerstat: A tool for measuring power consumption on linux,'' 2025, accessed: 2025-03-07. [Online]. Available: \url{https://github.com/ColinIanKing/powerstat}
\BIBentrySTDinterwordspacing

\end{thebibliography}

\end{document}